\documentclass[aps,pra,twocolumn,superscriptaddress,showpacs,showkeys,superscriptaddress]{revtex4-2}
\usepackage{dcolumn}    
\usepackage{ifpdf}
\usepackage{amssymb,lineno,amsfonts}
\usepackage{graphicx}   
\usepackage{dcolumn}    
\usepackage{bm}         
\usepackage{mathrsfs}
\usepackage{upgreek}
\usepackage{mathtools}
\usepackage{epstopdf}
\usepackage{setspace}
\usepackage{hyperref}
\usepackage{float}
\usepackage{amsmath}
\usepackage{placeins}
\usepackage{bbm}
\usepackage[normalem]{ulem}
\usepackage[utf8]{inputenc}
\usepackage[usenames,dvipsnames]{xcolor}
\usepackage[mathscr]{euscript}
\usepackage[]{qcircuit}
\usepackage[normalem]{ulem}

\definecolor{med-blue}{RGB}{25,25,112}
\hypersetup{colorlinks, linkcolor={red},citecolor={blue}, urlcolor={MidnightBlue}}

\newcommand{\ket}[1]{\vert{#1}\rangle}
\newcommand{\bra}[1]{\langle{#1}\vert}

\newcommand{\tr}{\mathrm{Tr}}

\begin{document}
	\title{Selective Wigner phase space tomography and \\ its application for studying quantum chaos}
	\author{Deepesh Khushwani }
\email{deepesh.khushwani@students.iiserpune.ac.in}
 \author{Priya Batra
 }
\email{priya.batra@students.iiserpune.ac.in}
\author{V. R. Krithika
 }
\email{krithika_vr@students.iiserpune.ac.in}
 \author{T. S. Mahesh 
}
\email{mahesh.ts@iiserpune.ac.in}
 	\affiliation{Physics Department and NMR Research Center, 
		Indian Institute of Science Education and Research, Pune 411008, India
	}
		
\begin{abstract}
			{
   The quasiprobability distribution of the discrete Wigner function provides a complete description of a quantum state and is, therefore, a useful alternative to the usual density matrix description.  Moreover, the  experimental quantum state tomography in discrete Wigner phase space can also be implemented.  We observe that for a certain class of states, such as harmonic states, the Wigner matrix is far more sparse compared to the density matrix in the computational basis.  Additionally, reading only a small part of the Wigner matrix may suffice to infer certain behavior of quantum dynamics.  In such cases, selective Wigner phase space tomography (SWPST) can be more efficient than the usual density matrix tomography (DMT).  Employing nuclear magnetic resonance methods on a three-qubit nuclear spin register, we  experimentally estimate Wigner matrices of various two-qubit quantum states.  As a specific example application of SWPST, we study the evolution of spin coherent states under the quantum chaotic kicked top model  and extract signatures of quantum-classical correspondence in the Wigner phase space.  
			}
	\end{abstract}
		

\maketitle

\section{Introduction}
 \label{Introduction}

Phase space is extensively used to study classical systems, where we can define an ensemble of particles with a probability density in the phase space \cite{goldstein2002classical}. However in quantum mechanics, the uncertainty principle and the wave nature of particles lead us to a quasiprobability distribution like the Wigner function \cite{quasiprobability,wigner}. 
The Wigner function  is a phase space function that keeps position and momentum on equal footing. Multiple studies have suggested that the continuous Wigner function can be used to explore the classical limit of quantum systems \cite{classlim1, classlim2}. The Wigner function has also been used in fields like signal processing and quantum optics \cite{Recentadv}. This led to several discrete Wigner phase space formulations, which have been used to investigate quantum resource theories \cite{negativeresource} and strongly correlated electronic systems \cite{dtwa, ctwa}.  One can also represent the states and gates of a quantum computer in the Wigner phase space and study properties of quantum algorithms \cite{Miquel}.  Wigner representations have been employed in the characterization and visualization of arbitrary operators \cite{leiner2017wigner}.   Direct quantum state tomography in Wigner phase space has also been described and experimentally demonstrated \cite{Miquel,Miquel2002}.

\begin{figure}
    \centering
    \includegraphics[width=0.49\textwidth]{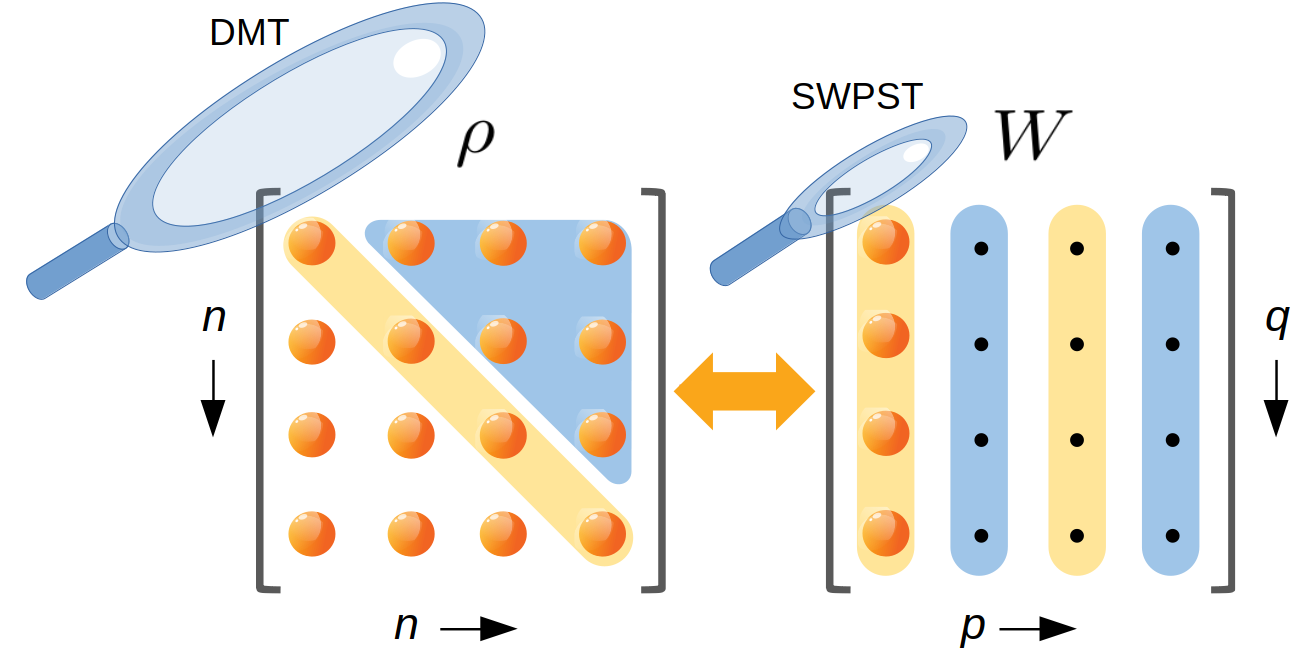}
    \caption{Illustrating the map between   density matrix $\rho$ in computational basis  to the Wigner phase space $W$.  It can take a nonlocalized and non-sparse matrix  to a localized and sparse matrix.  Therefore, selective Wigner phase space tomography (SWPST) is much more efficient than the usual density matrix tomography (DMT) for certain classes of states.
    In $W$, the even-numbered columns (starting from 0) correspond to the populations (diagonal elements of $\rho)$, and the odd-numbered columns correspond to the coherences (off-diagonal elements of $\rho$).}
    \label{fig:dm2wigner}
\end{figure}

The Wigner transform is analogous to the Fourier transform.  Just like a delocalized harmonic wave in the time axis becomes a delta function in the frequency axis, a class of quantum states that are delocalized in the computational basis can appear highly localized in either the position axis or momentum axis or both in the Wigner phase space.  Therefore, the corresponding discrete Wigner phase space can be far more sparse compared to the density matrix in the computational basis as illustrated in Fig. \ref{fig:dm2wigner}.  
Accordingly, here we propose selective Wigner phase space tomography (SWPST) that takes advantage of the higher sparsity of the Wigner matrix thereby  characterizing the  quantum states more efficiently than the usual density matrix tomography (DMT) in computational basis.  In general, complete $N$-dimensional DMT needs reading of $N^2$ complex elements, which may become prohibitively expensive for large $N$.  
On the other hand, partial DMT may suffice to know about certain interesting behaviours of quantum dynamics \cite{bonet2020nearly}.  Here we investigate quantum nonlinear systems via SWPST and  seek to extract signatures of quantum chaos. 

Classical nonlinear systems often exhibit chaotic dynamics, whose extreme sensitivity to the instantaneous location prevents long-term prediction of the phase space trajectory  \cite{strogatz2015nonlinear}.
The quantum-classical correspondence principle states that classical mechanics is reproduced by quantum mechanics in the limit of large quantum numbers \cite{correspondence}. 
Therefore, one expects some signatures of chaos in the quantum regime. However, this is not straightforward since the uncertainty principle means that trajectories are ill-defined in the quantum case. 
This correspondence between quantum and classical evolution in chaotic systems is the primary focus of quantum chaos, a subject with both fundamental and technological interest \cite{georgeot, chaudhury2009quantum, Grass}. 

Nuclear Magnetic Resonance (NMR) has been a successful test bed to implement quantum information processing tasks \cite{suter2008spins}. The nuclear spins controlled by radio-wave pulses offer long coherence times, precise controllability of spin dynamics, and efficient measurement of final states \cite{cory2000nmr}. Previously, the NMR architecture has been used to study various nonlinear quantum phenomena such as kicked top \cite{Krithika}, phase synchronization \cite{krithika2022observation}, bifurcations \cite{araujo2013classical} and  dynamical tunnelling \cite{krithika2023nmr}.

In this article, we look at quantum chaos in a two-qubit quantum kicked top (QKT) model implemented on a three-qubit system realized by three addressable spin-1/2 nuclei using NMR techniques. We apply radio frequency pulses as linear kicks, while the spin-spin interaction acts as the nonlinear evolution \cite{Krithika}. 
After each kick, we characterize the final state using SWPST, which allowed us to experimentally characterize quantum chaos more efficiently than DMT.

The article is organized as follows. In Sec. \ref{Theory}, we explain the mathematical backgrounds of the Wigner function, SWPST, and QKT. In Sec. \ref{Methodology}, we explain the NMR architecture and the experimental setup. Experimental results and discussion of the same are included in Sec. \ref{NMR-Results}. Finally, we conclude in Sec. \ref{Conclusions}.

\section{Theory}
\label{Theory}
\subsection{Wigner Function \label{sec:contwigner}}
The Wigner quasiprobability distribution function for a system described by the density matrix $\rho$ is defined as its Wigner-Weyl transform, which is given by \cite{wigner}

\begin{align} 
W(q,p) = \int \frac{d\lambda}{2\pi\hbar}e^{i\lambda p/ \hbar}
\bra{q-\lambda/2}{\rho}\ket{q+\lambda/2} ,
\end{align}
where $q$ and $p$ are conjugate position and momentum describing the phase space. 
The Wigner distribution function is characterized by the following properties \cite{Bertrand}:
\begin{enumerate}
    \item $W(q,p)$ is a quasiprobability distribution of real numbers that can also take negative values.  The negativity of the Wigner function at a particular point in phase space indicates that the corresponding quantum state cannot be described purely in terms of classical probabilities, and is thus a signature of quantum behaviour \cite{kenfack2004negativity}. 
    \item The fidelity overlap or the Frobenius inner product between states $\rho_1$ and $\rho_2$ with respective Wigner distributions $W_1(q,p)$ and $W_2(q,p)$ is given by \begin{equation}\label{eq:overlap}
    \tr[{\rho_1}{\rho_2}]=2\pi\hbar\int dq \:dp\: W_1(q,p)W_2(q,p).
    \end{equation} 
    \item The marginal probabilities along position and momentum are given by
    \begin{align}\label{eqn:probdens}
        \int\limits_{-\infty}^{+\infty} dp \:W(q,p) = \langle q|{\rho} |q\rangle ,~
        \int\limits_{-\infty}^{+\infty} dq \:W(q,p) = \langle p|{\rho} |p\rangle. 
    \end{align}
    For pure states, 
    \begin{align}
        \int\limits_{-\infty}^{+\infty} dp\: W(q,p) = |\psi(q)|^2  ,~  \int\limits_{-\infty}^{+\infty} dq\: W(q,p) = |\phi(p)|^2,
    \end{align}
    give probability density profiles along $q$ and $p$, respectively.
\end{enumerate}

It is important to note that the Wigner function is the only function with these defining properties, as compared to other quantum phase space measures \cite{Bertrand}. Accordingly, the Wigner function forms the basis for most definitions of finite dimensional quasiprobability distributions \cite{Ferrie}.

Given the Wigner function, we can determine the corresponding density matrix using the inverse mapping
\begin{equation}
    \rho=2\pi\hbar\int dq\:dp\: W(q,p)A(q,p),
\end{equation}
where $A(q,p)$ is called the phase space point operator, which is defined as \cite{WOOTTERS19871}
\begin{equation}
A(q,p)=
\frac{1}{(2\pi\hbar)^2}\int d\lambda \, d\lambda'\,e^{-i\lambda({P}-p)/\hbar+i\lambda'({Q}-q)/\hbar} .
\end{equation}
The phase space point operator $A(q,p)$ acts like a delta function at $(q,p)$ in the sense, given a density matrix $\rho$, the Wigner function can be obtained as the expectation value of $A(q,p)$ 
\begin{equation} 
    W(q,p)=\mathrm{Tr}[A(q,p)\rho].
\label{defofW}
\end{equation}

\subsection{Discrete Wigner Phase Space}
Multiple attempts have been made to generalize the continuous Wigner phase space definition to a finite-dimensional discrete Wigner phase space \cite{WOOTTERS19871,finitfelds,shalm2009squeezing,Ferrie}. The first generalization to satisfy all required properties was given by Wootters \cite{WOOTTERS19871} for a prime dimensional Hilbert space. Numerous others followed this, including spherical phase space \cite{shalm2009squeezing}, finite fields \cite{finitfelds} and even dimensional phase space \cite{Leonhardt}, each having different properties and applicability. We focus on the even-dimensional phase space proposed by Leonhardt \cite{Leonhardt}. It is a  popular definition of the discrete phase space, which can be used for any finite dimension of the Hilbert space.  Additionally, it has an easier visualization in the two-dimensional space \cite{Ferrie}.

The Leonhardt discretization of the Wigner function is achieved by transforming a Hilbert space of $N$ dimensions into a Wigner phase space of $2N$ dimensions \cite{Leonhardt}. This ensures that the three properties of the Wigner function described in Sec. \ref{sec:contwigner} are satisfied, as proven by Miquel et al. \cite{Miquel}. Here, we define  
\begin{align}
B_q &= \{|n\rangle, n=0,1,\cdots,N-1\} 
~~\mbox{and},
\nonumber \\
B_p&=\{|k\rangle, k=0,1,\cdots,N-1\}  
\end{align}
as the position and momentum basis, such that the momentum states $|k\rangle$ are given by the Fourier transform of the position basis,
\begin{equation}
    |k\rangle=\frac{1}{\sqrt{N}}\sum_n \exp(i2\pi nk/N)|n\rangle.
\end{equation}
In order to define the phase space point operators $A(q,p)$, we first setup the translation operators ${U}$, ${V}$ and ${R}$, given by 
\begin{align}
    {U}|n\rangle &=|n \oplus_N 1\rangle, ~~~ {U}|k\rangle = \exp(-i\: 2\pi k)|k\rangle,~ \nonumber  \\
    {V}|k\rangle &=|k \oplus_N 1\rangle, ~~~ {V}|n\rangle = \exp(i\: 2\pi n) |n\rangle,~\mbox{and} \nonumber \\
    {R}|n\rangle &=|N-n\rangle. 
\end{align}
Note that $U$ and $V$ are diagonal in momentum and position basis, respectively.
Then, $A(q,p)$ can be defined as 
\begin{equation}\label{defofA}
    A(q,p)=\frac{1}{2N}U^qRV^{-p}\exp(i\:\pi pq/N),
\end{equation}
where $q,p\in\{0,1,...2N-1\}$. Now, given a density matrix $\rho$, the corresponding discrete Wigner function $W(q,p)$ can be easily obtained by the expectation value of  $A(q,p)$ using Eq. \ref{defofW} \cite{Miquel}. 

The $2N$-dimensional Wigner phase space can be divided into four $N$-dimensional quadrants $(s_q,s_p)$ with $s_q,s_p\in \{0,1\}$ as below:
\begin{center}
$
\begin{array}{|c|c|}
\hline 
& \\
~(0,0)~ & ~(0,1)~  \\
& \\
\hline
& \\
~(1,0)~ & ~(1,1)~  \\
& \\
\hline
\end{array}.
$
\end{center}
In the following discussion, we denote the $(0,0)$ quadrant of $N\times N$ values as $G_N$ and the full phase space as $G_{2N}$. Although the phase space is defined in a $4N^2$ dimensional space, there are only $N^2$ independent phase space point operators in the quadrant $G_N$, and other three quadrants can be derived from these $N^2$ operators via \cite{Miquel}
\begin{align}\label{eq:linrel}
    A(q+s_qN,p+s_pN)&=A(q,p)(-1)^{s_pq+s_qp+s_qs_pN}, \nonumber \\
    \therefore~ W(q+s_qN,p+s_pN)&=W(q,p)(-1)^{s_pq+s_qp+s_qs_pN}.
\end{align} 
The marginal probabilities are given by \cite{Miquel}
\begin{align}
    \sum_pW(q_0,p)&=\langle  q_0/2|{\rho}|q_0/2\rangle~~\mbox{and}
    \nonumber \\
    \sum_q W(q,p_0)&=\langle  p_0/2|{\rho}|p_0/2\rangle.
\end{align}
From the symmetry relations Eq. \ref{eq:linrel}, it turns out that, for an even $q_0$, the sum of elements over the line $q=q_0$ gives the probability of obtaining the state $|n=q_0/2\rangle$. For an odd $q_0$, the sum of elements over the line $q=q_0$ simply vanishes. Therefore, for classical mixtures of the computational basis states, we find that all the odd columns have zero values, while for quantum superposition states, we find them having non-zero values. Thus, the odd columns signify the coherences in the density matrices, and the even columns signify the populations as illustrated in Fig. \ref{fig:dm2wigner}. 

Given two discrete Wigner matrices $W_1$ and $W_2$ corresponding to the density matrices $\rho_1$ and $\rho_2$, the fidelity or the Frobenius inner product of the two states is given as
\begin{align}
F = \tr[\rho_1\rho_2] = 
4N\sum_{(q,p) \in G_{N}}
W_1(q,p) W_2(q,p),
\label{eq:fidelity}
\end{align}
and infidelity as $1-F$.

\subsection{Selective Wigner Phase Space Tomography (SWPST)}
\begin{figure}
    \centering 
\includegraphics[clip=,trim=0cm 0cm 0cm 0cm,width=7.5cm]{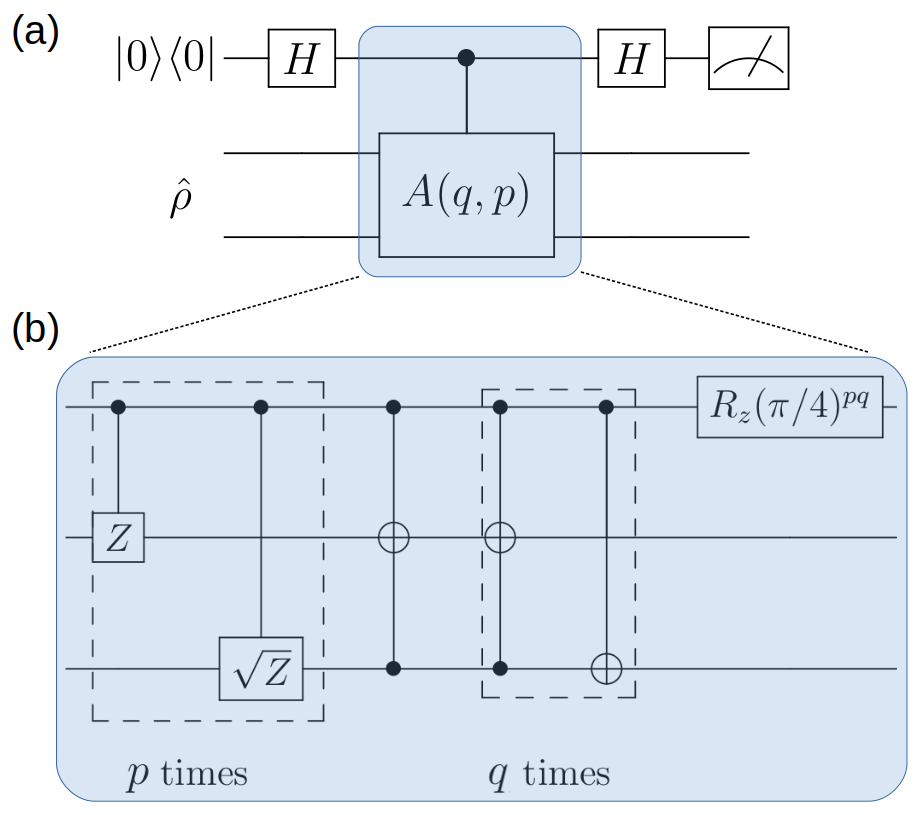}
    \caption{(a) Quantum circuit for the measurement of $W(q,p)$ using an ancillary qubit (top qubit) and an interferometric approach to extract the expectation value of Eq. \ref{defofW}. (b) Circuit for realizing the point space operator $A(q,p)$ for a 2+1 qubit register.}
    \label{fig:circuit}
\end{figure}

\begin{figure}
    \centering
\includegraphics[clip=,trim=1cm 1.5cm 1cm 1cm,width=8.5cm]{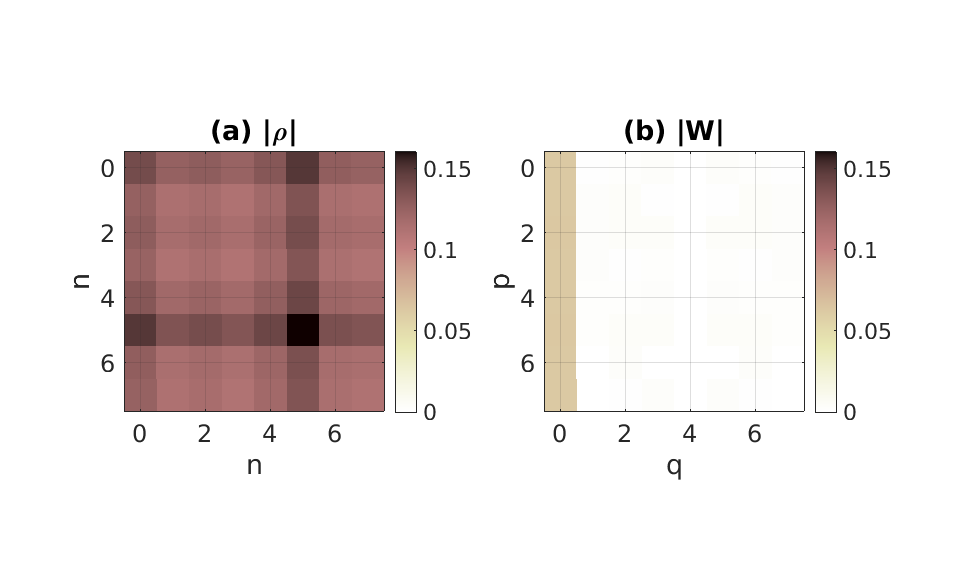} \\
     \vspace*{-0.5cm}
\includegraphics[clip=,trim=0cm 0cm 0cm 0cm,width=8.5cm]{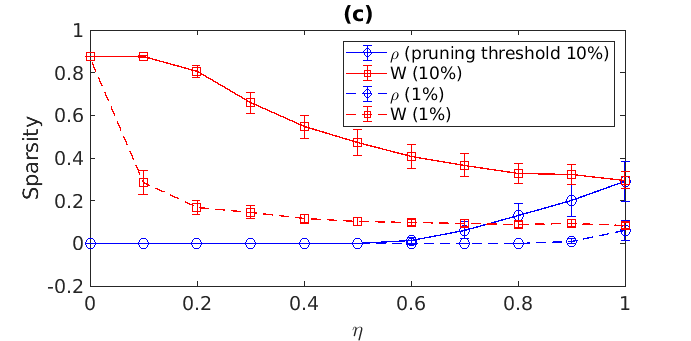}
\\
\vspace*{-0.0cm}
\includegraphics[clip=,trim=0cm 0cm 0cm 0cm,width=8.5cm]{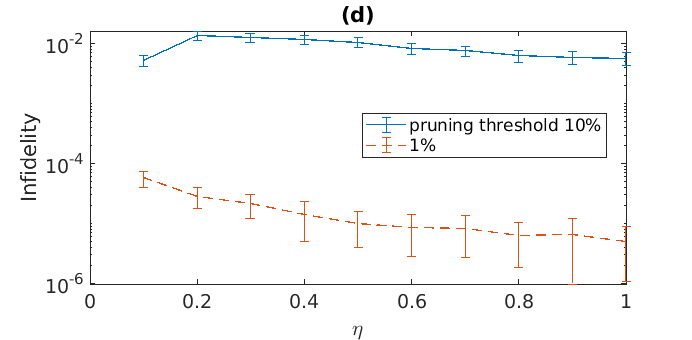}
\caption{(a,b) The density matrix $\rho$ in computational  basis $B_q$ and the Wigner matrix $W_{\{G_N\}}(q,p)$ in the $(B_q,B_p)$ basis for the 3-qubit randomized harmonic state $\ket{\psi_0(\eta)}$ with $\eta=0.1$ (see Eq. \ref{eq:randomharmonicstate}).  The matrix in (a) is not sparse, while in (b)  $87.4 \% $ of elements have magnitudes below 10 \%  and $28 \pm 5$ \% of elements have magnitudes below 1 \% of the highest elements (left-most column in $W$). (c) Sparsity yielded with pruning threshold 10 \% (solid lines) and 1 \% (dashed lines) in $\rho$ (circles) and $W$ (squares) matrices corresponding to $\ket{\psi_0(\eta)}$ versus the randomization parameter $\eta$. (d) The infidelities between unpruned Wigner matrix with itself after pruning at threshold 10\% (solid line) and 1\% (dashed lines).  }
    \label{fig:rhovsW}
\end{figure}

Wigner phase space tomography is carried out by measuring the values of $W(q,p)$ as the expectation value of Hermitian operators $A(q,p)$ (Eq. \ref{defofW}).  The quantum circuit in Fig. \ref{fig:circuit} employs an interferometric approach to extract the Wigner elements with the help of  an ancillary qubit \cite{Miquel}.
Naively, it may appear computationally more expensive to carry out the Wigner phase space tomography.  However, we observe that it is often not necessary to read the complete Wigner matrix.  Consider, for example, the harmonic states of the form 
\begin{align}
\ket{\psi_j} = \mbox{QFT}(\ket{j}) = 
\frac{1}{\sqrt{N}}\sum_{n=0}^{N-1} \exp(i2\pi nj) \ket{n},
\label{eq:harmonicstate}
\end{align}
obtained by quantum Fourier transform (QFT).
This state has equal amplitudes over all the computational basis states  and therefore $\rho$ is completely nonsparse (see Fig. \ref{fig:rhovsW} (a)).  However, as seen in Fig. \ref{fig:rhovsW} (b), its $W(q,p)$ has only one non-zero column corresponding to the $\ket{p=j}$ vector.  Thus, $W(q,p)$ has a high sparsity of $(N-1)/N$.  To show the robustness of sparsity we multiply each amplitude by a random number $r_n \in [1-\eta,1+\eta]$ to obtain the randomized harmonic state
\begin{align}
\ket{\psi_j(\eta)} = 
A\sum_{n=}^{N-1} r_n \exp(i2\pi nj) \ket{n},
\label{eq:randomharmonicstate}
\end{align}
where $\eta\in[0,1]$  and $A$ is the normalization factor.
Even when amplitudes are randomized, the harmonic states remain highly localized in $W$.  Pruning of the localized $W$ matrix by nullifying weak elements below a certain threshold value, we obtain highly sparse matrix as shown in Fig. \ref{fig:rhovsW} (c).  Note that here pruning does not significantly affect the quantum state information as indicated by the low infidelity values plotted in Fig. \ref{fig:rhovsW} (d).  
Thus, the high sparsity of pruned $W$ matrix allows a partial tomography of only residual elements thereby leading us to SWPST which is more efficient in state characterization compared to complete DMT.  In the following, we shall describe a specific example which we have experimentally investigated.

\subsection{Quantum Chaos in Quantum Kicked Top}
Signatures of quantum chaos have been studied using various quantities like quantum correlations such as entanglement and quantum discord \cite{chaoscorr,Unentch}, level statistics \cite{levelstat1} and dynamics of open quantum systems undergoing continuous measurement \cite{oqs1}. Recently, a machine learning protocol to efficiently compute QKT phase space in terms of quantum discord has also been proposed \cite{batra2021efficient}.

\begin{figure*}
    \centering
    \includegraphics[clip=,trim=4.5cm 0cm 4.5cm 0cm,width=0.99\textwidth] {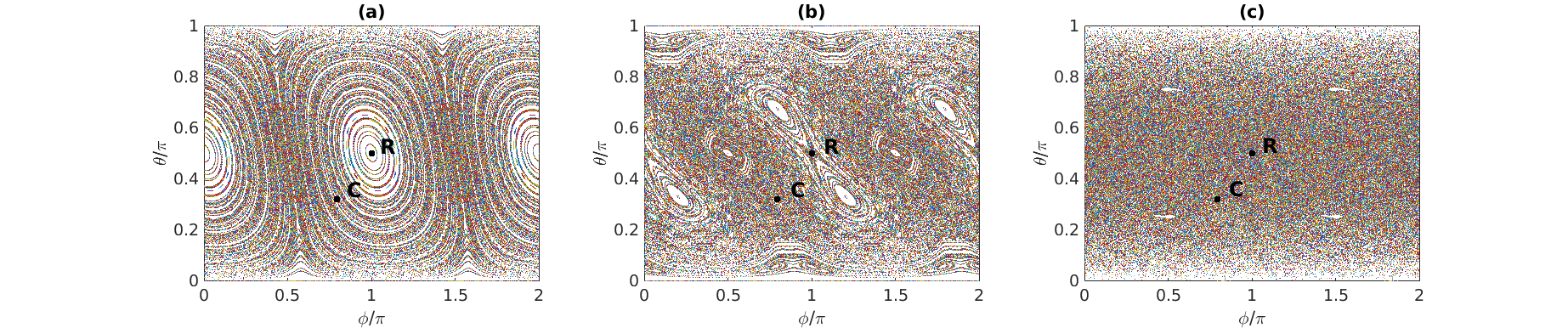} \caption{Phase space of QKT in the classical limit showing (a) periodic trajectories for low values of chaoticity parameter ($k=0.5$), (b) periodic islands separated by chaotic sea for intermediate chaoticity values ($k=2.5$), and (c) completely chaotic phase space for larger chaoticity values ($k=2\pi+2.5$). Two phase space regions are chosen for detailed analysis: R (from a regular region) and C (from a chaotic region). }
    \label{fig:qktcps}
\end{figure*}

In this work, we study quantum chaos from the phase space perspective using Wigner functions. The QKT model of spin systems is one of the most studied models of quantum chaos \cite{Entassigofchaos, QKT2, Unentch}. The QKT Hamiltonian for finite kick width is given as (with $\hbar=1$) \cite{Haake1987}, 
	\begin{align}
	H_{\mathrm{QKT}} = \begin{cases}
	H_{\mathrm{kick}} = \frac{\pi}{2\Delta}J_x,~{\mathrm{for~}}t\in \left[n\tau-\frac{\Delta}{2},n\tau+\frac{\Delta}{2}\right]\\
	H_{\mathrm{NL}} = \frac{k}{2j\tau}J_z^2, ~{\mathrm{otherwise.}}
	\end{cases} 
	\label{Hqkt}
	\end{align}
Here $J_x$, $J_y$ and $J_z$ are the components of the angular momentum operator $\textbf{J}$ of a spin-$j$ system, $\Delta$ is the kick-duration, $\tau$ is the time between kicks,  and $k$ is the chaoticity parameter. 
The Hamiltonian leads to a unitary evolution such that 
\begin{align}
    &U_\mathrm{QKT}=U_\mathrm{NL}U_\mathrm{kick}, ~~\mbox{where,}
    \nonumber \\
    &U_\mathrm{NL}=e^{-ikJ_z^2/2j}  ~~ \mbox{and}~~ U_\mathrm{kick}=e^{-i(\pi/2) J_x}.
\end{align}

In the classical limit, this model results in regular regions of phase space with periodic trajectories for small values of $k$, while chaotic regions gradually emerge for larger values of $k$, as shown in Fig. \ref{fig:qktcps}. 
To study QKT, one conventionally  uses spin coherent states $|\theta,\phi\rangle$ as these are minimum uncertainty states and are regarded closest to classical states \cite{scsclassical}. 
Thus, for any initial spin coherent state,  the dynamics are highly regular for low values of $k$, while,  depending on the initial conditions, one expects a mixed behaviour for higher values of $k$.  

The QKT model is also experimentally realizable with qubits since the QKT dynamics for a spin-$j$ system can be mapped to the symmetric subspace of $2j$ qubits \cite{Entassigofchaos}.  QKT model with two NMR qubits has been experimentally studied  in Ref. \cite{Krithika}.  Recently, dynamical tunnelling of QKT between two regular regions has also been studied \cite{krithika2023nmr}.

\section{NMR Methodology}
\label{Methodology}
\subsection{Spin system and the Hamiltonian}
\begin{figure}
    \centering \includegraphics[clip=,trim=0cm 0cm 0cm 0cm,width=8cm]{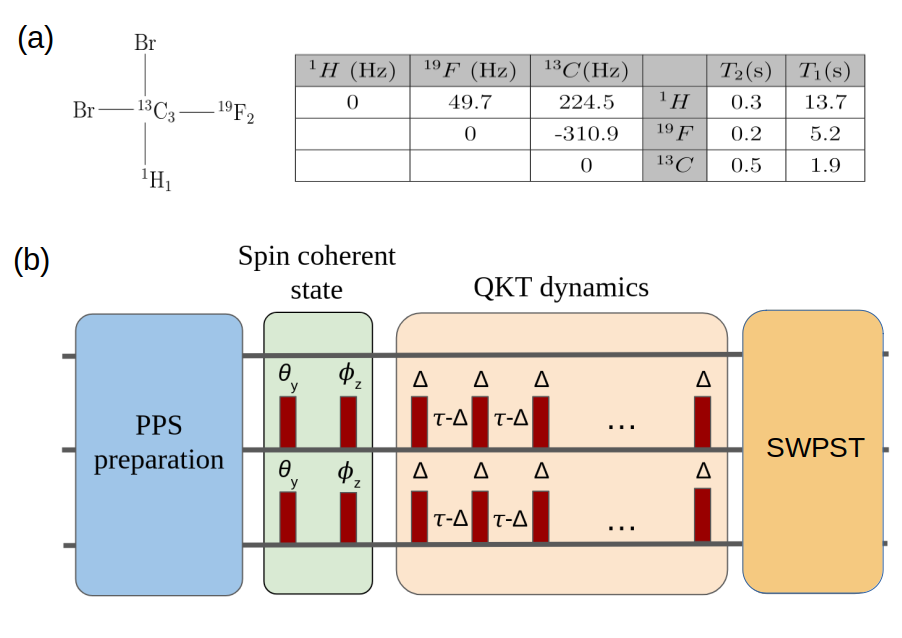} 
    \caption{(a) Molecular structure of DBFM, resonance offsets (diagonal elements), coupling strengths (off-diagonal elements), and $T_1,~T_2$ relaxation time constants.  (b) The NMR pulse sequence for studying QKT using SWPST.}
    \label{fig:nmrpul}
\end{figure}

Our 3-qubit quantum register involves $^{1}$H, $^{13}$C and $^{19}$F nuclei  of dibromofluoromethane (DBFM)  dissolved in deuterated acetone. Fig. \ref{fig:nmrpul} (a) shows the molecular structure, Hamiltonian parameters, and the relaxation time constants of DBFM.
All experiments were performed on a 500 MHz Bruker NMR spectrometer operating at a magnetic field $B_0 = 11.75$ T at an ambient temperature of 300K.
The Zeeman interaction leads to an energy gap $\hbar \omega_i = \hbar \gamma_i B_0$  between the spin states corresponding to $m = \pm 1/2$ where $\gamma_i$ is the gyromagnetic ratio of the nuclear isotope and $\omega_i=\gamma_i B_0$ is the Larmor frequency \cite{levitt2013spin}. The spins also interact with one another through scalar coupling  $J_{ij}$, which is mediated by covalent bonds. 
Under high-field, high-temperature, and weak-coupling approximation, the NMR Hamiltonian in a frame co-rotating with the individual radio-frequency (RF) carriers of each nuclear isotope is of the form
\begin{align}
H_{\mathrm{NMR}} &= H_\mathrm{\nu} + H_{J} + H_{\mathrm{RF}},
~~\mbox{where}
\nonumber \\
H_\mathrm{\nu} & = -2\pi\sum_{i=1}^{N}  \nu_i {I}_{iz}, 
\nonumber \\
H_J &= 2\pi\sum_{i,j>i}^{N} J_{ij} {I}_{iz} {I}_{jz},
~~\mbox{and}
\nonumber \\
H_{\mathrm{RF}} &= 2\pi \sum_{i=1}^N \Omega_{i}(t) \left( \cos\Phi_i(t) {I}_{ix} + \sin\Phi_i(t) {I}_{iy} \right).
\label{intH1}
\end{align}
Here $\nu_i$ are the resonance offsets, $\Omega_i(t)$ are the RF amplitudes, $\Phi_i(t)$ are RF phases, and $I_{i\alpha}$ are the spin angular momentum operators.  Under on-resonant condition ($\nu_i=0$), an RF pulse of duration $\Delta$ with a constant amplitude $\Omega_i$ and phase $\Phi_i = 0$ produces an $i$th spin rotation along $x$ by $\theta_x = 2\pi \Omega_i \Delta$.  Similarly, other rotations can be generated by precisely controlling the amplitudes and phases of the RF fields.

The important steps involved in the experiments are illustrated in Fig. \ref{fig:nmrpul} (b).  In the following, we describe these steps in detail.

\subsection{Initialization, Dynamics, and  Readout}
Under high-temperature and high-field approximation, the NMR initial states are of the form 
\begin{equation}
\rho_0=\mathbbm{1}/2^N+ \sum_i  \epsilon_i {I}_{iz},
\end{equation}
where $\epsilon_i=\gamma_iB_0/(2^nk_BT) \sim 10^{-5}$ are known as the purity factors. Here, the identity gives the uniform background population that is invariant under unitary transformation, while the second part is the deviation density matrix that captures all the interesting dynamics. Realizing a pure NMR state $\rho_\mathrm{pure}$ requires impractically low temperatures and extremely high Zeeman fields. Therefore, it is customary to prepare a pseudopure state (PPS) that is of the form \cite{cory1997ensemble,knill1998effective}
\begin{equation}
\rho_\mathrm{PPS} = (1-\epsilon)\mathbbm{1}/2^N+ \epsilon \rho_\mathrm{pure}.
\end{equation}
With an effective purity of $\epsilon$, the
PPS  captures the essential dynamics of a pure state.  The NMR pulse sequence for preparing the PPS is similar to the one in Ref. \cite{krithika2021observation}.

After preparing the $|000\rangle$ PPS, a $\theta_y$ pulse followed by a $\phi_z$ pulse on the second and third qubits initializes the qubits to a spin coherent state 
\begin{equation}\label{scs}
    \rho_{\theta,\phi} = \ket{0}\bra{0}\otimes\ket{\theta,\phi}\bra{\theta,\phi}\otimes\ket{\theta,\phi}\bra{\theta,\phi} , 
\end{equation}
where $\ket{\theta,\phi}=\cos(\theta/2)\ket{0}+e^{i\phi}\sin(\theta/2)\ket{1}$ is the Bloch vector for each system qubit
(see Fig. \ref{fig:nmrpul} (b)). This initialization is equivalent to the classical initialization in the $(\theta,\phi)$ point on the classical phase space. 

We used DBFM as the system, wherein $^1$H was used as the ancillary qubit (see Fig. \ref{fig:nmrpul} (a)).
To describe the experimental study of the QKT Hamiltonian, we first note that the nonlinear $z$ angular momentum term \begin{equation}
    J_z^2 = (I_{z1}+I_{z2})^2 = \mathbf{I}/4+\mathbf{I}/4+2I_{z1}I_{z2}.
\end{equation} 
Dropping the identities, since they only give rise to a global phase, we note that the interesting dynamics of the nonlinear Hamiltonian can be realized using the bilinear term. Therefore, the nonlinear interaction of QKT is realized by evolving spin-spin interaction Hamiltonian $H_J$  
for time $\tau$ such that $\tau=k/2\pi J_{12}$ (see Fig. \ref{fig:nmrpul} (b)). The kicks, given by the Hamiltonian
\begin{equation}
    H_\mathrm{RF}=\frac{\pi}{2\Delta_1} I_{1x}+\frac{\pi}{2\Delta_2}I_{2x}, 
\end{equation}
are applied using RF pulses on both the qubits with duration $\Delta_i = 1/(4\Omega_i)$. This corresponds to a $\pi/2$ pulse in the $x$-direction. 

The readout circuit for SWPST is shown in Fig. \ref{fig:circuit} (b).  We realized the Toffoli gate by designing the amplitude and phase-modulated RF pulses with the help of  the Gradient Ascent Pulse Engineering (GRAPE) method \cite{KHANEJA2005}. The numerical fidelity of the GRAPE pulse was above 0.99 after averaging over $\pm 10\%$ inhomogeneity of RF fields.  All other gates of Fig. \ref{fig:circuit} (b) were realized by rectangular pulses and evolutions under spin-spin interactions.

\section{Results}
\label{NMR-Results}
\subsection{Wigner phase space tomography of some standard 2-qubit states}
\begin{figure}
\centering
\small{(a)} \\
\includegraphics[clip=,trim=3cm 0cm 1.5cm 0cm,width=9cm]{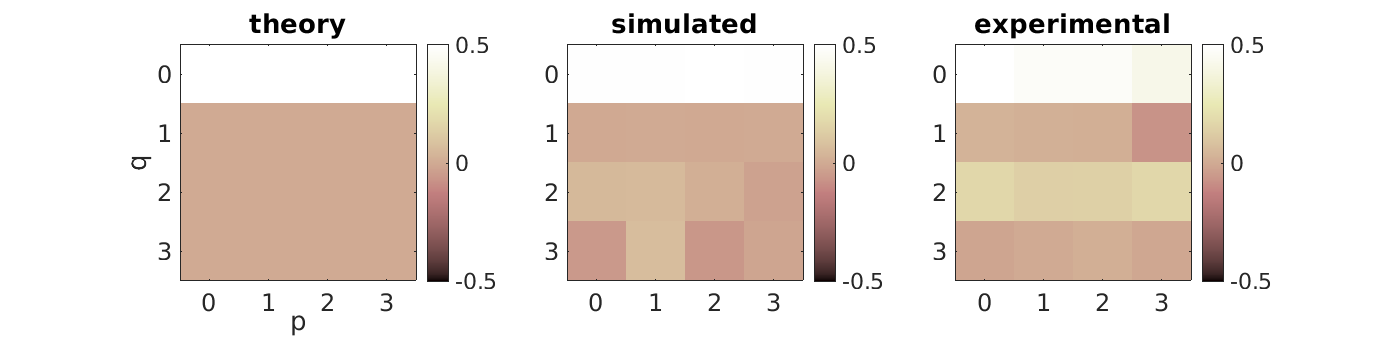} 
\\
\small{(b)} \\
\includegraphics[clip=,trim=3cm 0cm 1.5cm 0cm,width=9cm]{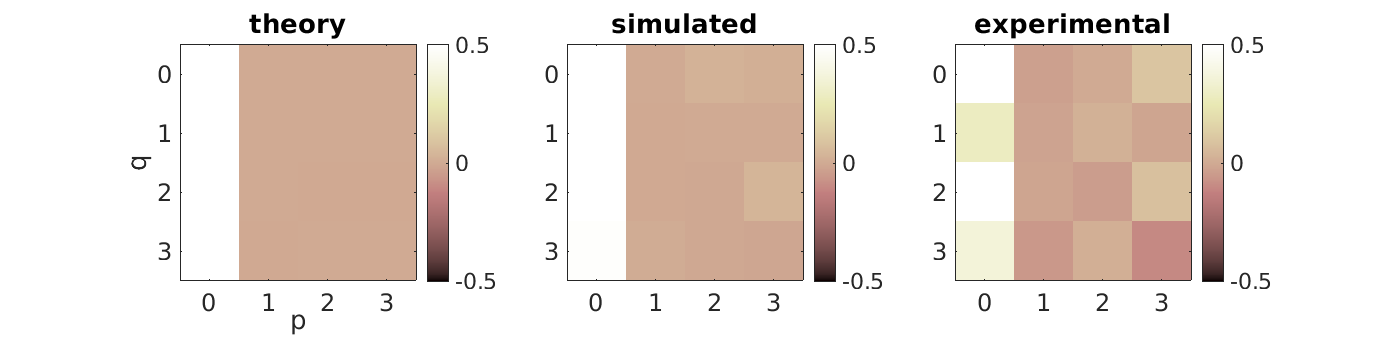}
\\
\small{(c)} \\
\includegraphics[clip=,trim=3cm 0cm 1.5cm 0cm,width=9cm]{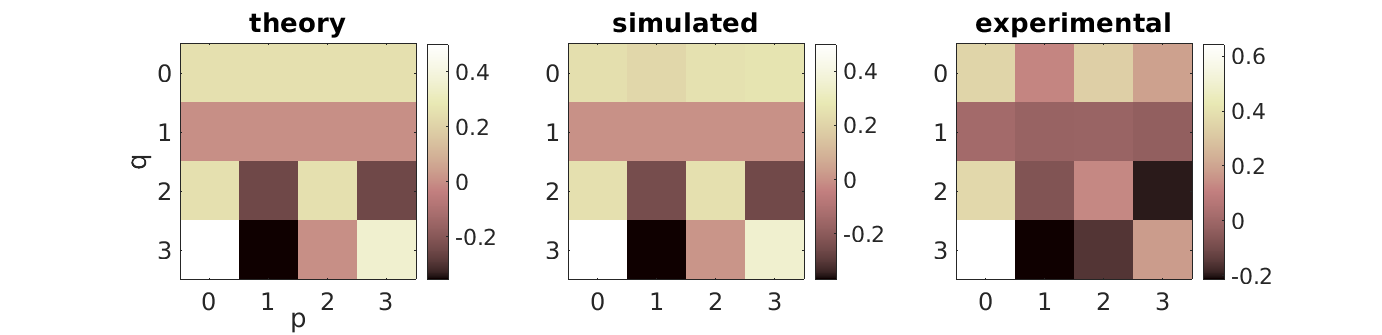}
\caption{Heat maps representing theoretical, simulated (using GRAPE pulses with gate-fidelity of 0.99), and experimental two-qubit Wigner matrices for (a) $|00\rangle$ $(0.94)$, (b)  $|++\rangle$ $(0.94)$, (c) Bell state $(|00\rangle+|11\rangle)/\sqrt{2}$ $(0.90)$, where the values in parenthesis are fidelities of experimental states with theoretical states.  
}
\label{fig:expwigtom}
\end{figure}

\begin{figure*}
  \includegraphics[clip=,trim=0cm 10cm 0cm 0cm,width=0.85\textwidth]{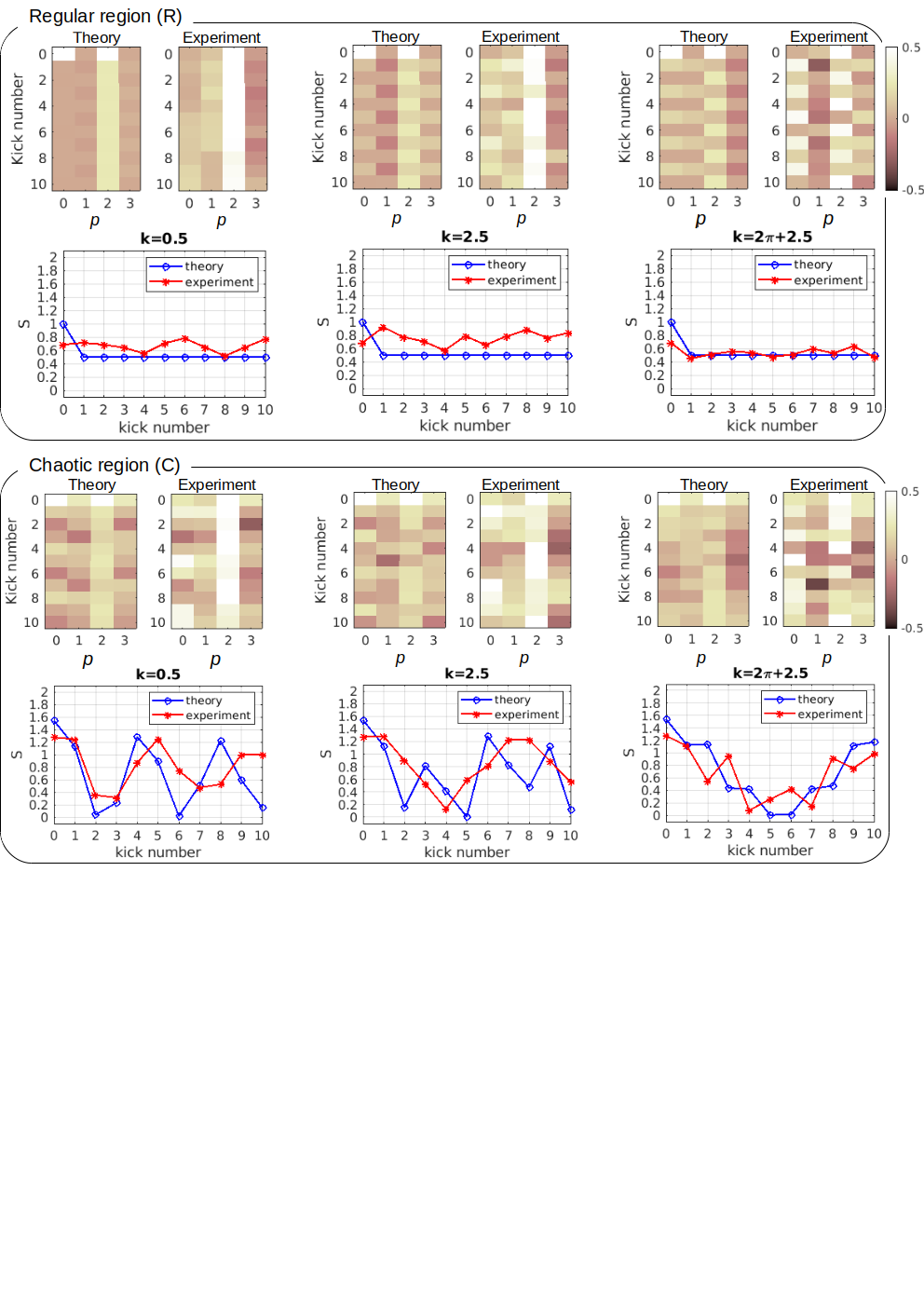}
\caption{SWPST results for QKT dynamics starting from each of the two regions marked in Fig. \ref{fig:qktcps} at different chaoticity values ($k$) as indicated. Each row of the heat maps represents the first row of the Wigner matrix summing to $S$ ($S = \sum_{p\in\{0,1,2,3\}} W(q=0,p)$), and transforming under the nonlinear evolution.}
    \label{fig:qktresults}
\end{figure*}

We first use the circuit in Fig. \ref{fig:circuit} (a) to experimentally carry out the complete Wigner phase space tomography of certain standard two-qubit states, specifically product states (a) $|00\rangle$, (b)  $|++\rangle$, and the Bell state (c) $(|00\rangle+|11\rangle)/\sqrt{2}$. 
Fig. \ref{fig:expwigtom} shows the theoretical (with ideal operators), simulated (with GRAPE pulses), and experimentally obtained Wigner matrices.  The experimental fidelities of product states with the theoretically expected states were $0.94$  while that for the Bell state was $0.90$.  Somewhat lower values of fidelities can be attributed to both  preparation errors and  tomography errors.  The errors are predominantly due to experimental imperfections like RF inhomogeneity, nonlinearities of RF amplifiers, static field inhomogeneity, and  decoherence effects.  Note that both the product states discussed above have highly sparse Wigner matrices. In Fig. \ref{fig:expwigtom} (a), only the first row is nonzero, while in Fig. \ref{fig:expwigtom} (b), only the first column is nonzero.  Such type of states are efficiently characterized by SWPST.

In the following, we describe a situation where SWPST can be advantageous even for nonsparse matrices.

\subsection{Signatures of quantum chaos}
Now, we describe the results of SWPST used to detect signatures of chaos in the QKT model. As described in Fig. \ref{fig:qktcps}, we selected two phase space points $(\theta,\phi) \in \{R\equiv (\pi/2,\pi), ~C\equiv (1.0,2.5)\}$, which respectively correspond to a regular region and a chaotic region.  As depicted in the pulse diagram shown in Fig. \ref{fig:nmrpul} (b), we drive each spin-coherent state through QKT evolution with each of the three chaoticity values ($k\in \{0.5,~2.5,2\pi+2.5\}$, and finally perform SWPST.

In these experiments, SWPST involves reading only the first row of the Wigner matrix $W(q=0,p\in\{0,1,2,3\})$ after each kick, and tracking the sum
\begin{align}
S = \sum_{p\in\{0,1,2,3\}} W(0,p).
\end{align}
Based on numerical comparisons, we found this region of the Wigner phase space to be sensitive to the extent of quantum chaos.
The results are shown in Fig. \ref{fig:qktresults}.  For the initial spin coherent state $R\equiv (\pi/2,\pi)$ corresponding to the central regular region of the classical phase space (see Fig. \ref{fig:qktcps}), the evolution remains confined within a small region of the phase space even for high chaoticity values.  Accordingly, the scalar quantity $S$, after an initial jump, remains mostly static.  Such a behaviour is the characteristic of nonchaotic dynamics.
In contrast, for the second spin coherent initial state $C\equiv (1.0,2.5)$, corresponding to a chaotic region, the evolution explores a larger area of phase space that increases with the chaoticity parameter.  Accordingly, the Wigner elements and the scalar quantity $S$ wildly oscillate, indicating the expected chaotic behaviour.  

In short, reading only a small portion of the Wigner matrix by SWPST enables us to distinguish regular behavior from chaotic dynamics.  Furthermore, the mismatch between the theory and experiments is more pronounced in the chaotic dynamics compared to the regular dynamics.  Similar observations have practical implications in quantum control and recently have gained significant attention \cite{Krithika,berke2022transmon}.

\section{Conclusions}\label{Conclusions}
Efficient characterization of quantum systems is at the heart of upcoming quantum technologies.  Density matrix tomography (DMT), for instance, involves the estimation of all the density matrix elements.  While it provides a complete description of the quantum state, its complexity scales exponentially with the size of the system and quickly becomes impracticable for larger systems.  
Earlier, there have been several studies on Wigner phase tomography and their experimental demonstrations.  
In this work, we proposed selective Wigner phase space tomography (SWPST) and described its advantage using numerical analysis of the sparsity of Wigner matrices. We also experimentally demonstrated the advantage of SWPST in probing quantum systems using a three-qubit NMR quantum register.

Wigner transform is analogous to the Fourier transform, which takes a highly distributed waveform, such as a harmonic function in time space, to a highly localized distribution, such as a delta function, which allows an efficient reading of frequency.  In the same way, a harmonic quantum state in the computational basis is spread out throughout the density matrix, rendering DMT a hard task.  The same harmonic state appears as a highly sparse matrix in the Wigner phase space, allowing us to extract information with only a partial reading of the Wigner matrix.  
We numerically showed a 3-qubit harmonic state to have a high degree of sparsity of over 87 \% in the Wigner matrix, while its density matrix being completely non-sparse.  Even after gradual amplitude randomization the harmonic states retained a sparsity of up to about 30 \% in their Wigner matrices.
Thus, we indicated the existence of a class of quantum states for which SWPST can be far more efficient than DMT in the computational basis.  
As proof of principle demonstration of Wigner tomography, we experimentally estimated complete Wigner matrices of two product states and one entangled state of a two-qubit system with the help of a third ancillary qubit.

We also described the applicability of SWPST in studying chaotic quantum systems. Here, we used the quantum kicked top model, which involves a repeated application of linear kicks and nonlinear evolutions.  By reading only a small part of the Wigner matrix we could distinguish regular dynamics from chaotic dynamics. 
Such efficient characterization protocols can be helpful in exploring the effect of chaos in quantum computers and will assist in achieving quantum control and high-fidelity quantum computation in the presence of quantum chaos \cite{berke2022transmon}. 

We believe that the Wigner function formalism can also help in exploring the semi-classical approximation and the quantum to classical boundary.
It may also be helpful in studying exotic phenomena like quantum scars \cite{scars} and dynamical tunnelling \cite{tunneling, krithika2023nmr} that occur as a result of chaos.

We also envision the possibility of implementing an approximate Wigner phase space tomography.  For instance, approximate implementation of quantum Fourier transform often involves dropping high-depth control gates \cite{barenco1996approximate}.  Along the same lines, can we construct an approximate Wigner phase space tomography by a depth-optimized execution of the phase space point operator?  Such a construct  would be highly useful for experimental implementations and similar questions can be addressed in future works.

\section*{Acknowledgments}
D. K. acknowledges technical guidance from Mr. Vishal Varma, Dr. Sandeep Mishra and Mr. Nitin Dalvi on NMR operations. T.S.M. acknowledges funding from
DST/ICPS/QuST/2019/Q67. We thank the National Mission on Interdisciplinary Cyber Physical Systems for funding from the DST, Government of India through the I-HUB Quantum Technology Foundation, IISER-Pune. P.B. gratefully acknowledges the Prime Minister's Research Fellowship of Govt. of India for financial support.

\bibliography{references}

\end{document}